\documentclass{ws-procs9x6}

\begin{document}
\title{Photoproduction of $K^{*+}\Lambda$ and $K^+\Sigma(1385)$ in the reaction $\gamma \lowercase{p} \rightarrow K^+ \Lambda \pi^0 $ at Jefferson Lab\footnote{\uppercase{T}his work is supported by \uppercase{DOE} under contract \uppercase{DE-AC05-84ER40150}}}
\author{L. Guo and D.~P. Weygand\\ For the CLAS Collaboration}
\address{Jefferson Lab, \\
12000 Jefferson Ave., 
Newport News, VA 23606, USA\\
E-mail: lguo@jlab.org}

\maketitle

\abstracts{
The search for missing nucleon resonances using coupled channel analysis has mostly been concentrated on $N\pi$ and $KY$ channels, while the contributions of $K^*Y$ and $KY^*$ channels have not been investigated thoroughly mostly due to the lack of data.  With an integrated luminosity of about 75 $pb^{-1}$, the photoproduction data using a proton target recently collected by the CLAS Collaboration at Jefferson Lab with a photon energy range of 1.5-3.8~GeV provided large statistics for the study of light hyperon photoproduction through exclusive reactions. The reaction $\gamma p \rightarrow K^+ \Lambda \pi^0$  has been investigated. Preliminary results of the $K^{*+}\Lambda$ and $K^+\Sigma(1385)$ cross sections are not negligible compared with the $KY$ channels. The $\Lambda \pi^0$ invariant mass spectrum is dominated by the $\Sigma(1385)$ signal and no significant structure was found around the $\Sigma(1480)$ region. 
}

\section{Introduction}
Presently, coupled channel analysis has been mostly performed for $N^*$ physics by including pion, eta, and kaon production. However, it should be pointed out that $K^*Y$ and $KY^*$ production cannot be ignored. A recent study of $N^*\rightarrow K^*Y / KY^*$ by Capstick and Roberts~\cite{Capstick98}, using a quark-pair creation model, has predicted that a few low-lying negative-parity states could couple strongly to the $K^*\Lambda$ channel. The well established $N[\frac{7}{2}^-]_1(2190)$ from pion production data should also be clearly visible in photon reactions, while the two-star state $N[\frac{3}{2}^-]_1(2080)$ and the weakly established state $N[\frac{1}{2}^-]_1(2090)$ are predicted to couple to $K\Lambda$ and $K^*\Lambda$ with similar strength. As for the $KY^*$ channel , in particular the $K\Sigma(1385)$, the nominal threshold is lower, and several $N^*$ states are predicted to couple strongly to $K\Sigma(1385)$. These states include $N[\frac{1}{2}^-]_5(2070)$ (established in pion production), $\Delta[\frac{3}{2}^-]_3(2145)$, and the relatively lighter predicted states such as  $N[\frac{5}{2}^+]_2(1980)$ and $\Delta[\frac{3}{2}^+]_3(1985)$. The new CLAS data (g11) represents a total luminosity of about $75~pb^{-1}$ with a tagged photon energy range of 1.5-3.8~$GeV$~\cite{CLAS}$^{,}$~\cite{TAGGER}, making it possible to study the photoproduction of  $K^*Y$ and $KY^*$ in many channels which previously lacked statistics. 
\section{Basic data features}
\begin{figure}[htbp]
\begin{center}
\includegraphics[width=4.5in, height=2.1in]{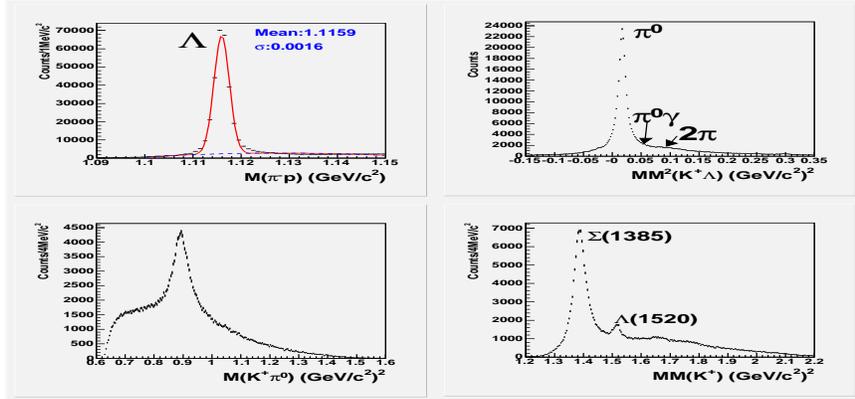}
\caption{Top Left: Invariant mass spectrum of $p\pi^-$ system. Top right: Missing mass squared spectrum off of the $K^+\Lambda$ system. Bottom left: Invariant mass spectrum of the $K^+\pi^0$ system.  Bottom right: $K^+$ missing  mass spectrum. }
\label{yyy}
\end{center}
\end{figure}

\begin{figure}[htbp]
\begin{center}
\includegraphics[width=4.5in, height=1.3in]{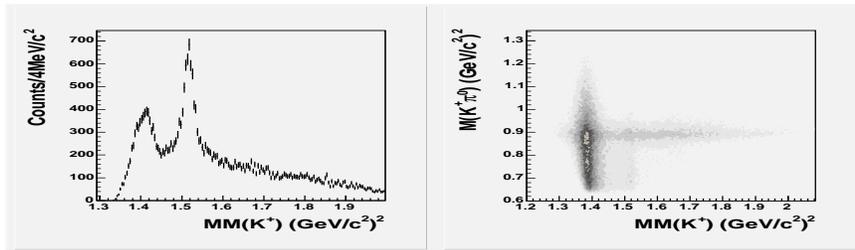}
\caption{Left: $K^+$ missing  mass spectrum  for the $\pi^0\gamma$ region indicated by Fig.~\ref{yyy}(Top right); Right: Invariant mass of the $K^+\pi^0$ system vs. $K^+$ missing mass.}
\end{center}
\label{kstar}
\end{figure}

The reaction of $\gamma p \rightarrow K^+ \Lambda \pi^0$ has been investigated by selecting events with the $K^+ \pi^-$ and proton being detected by the CLAS spectrometer. The $\Lambda$ is identified from the invariant mass of the $p\pi^-$ system, while the $\pi^0$ is reconstructed from the $K^+\Lambda$ missing four momentum (Fig.~\ref{yyy}, top). Overall, the data is dominated by $K^*\Lambda$ and $K^+\Sigma(1385)$ production (Fig.~\ref{yyy}, bottom). Recently, there has been some renewed interest in a possible $\Sigma(1480)$ state, with the latest observation reported by the COSY collaboration in the reaction $p p \rightarrow p K^+ Y^{0*}$~\cite{ZYCHOR}. No significant signal was found in the $\Lambda\pi^0$ invariant mass spectrum for any range of photon energy and other kinematic variables. However, direct comparison is difficult due to the different production mechanisms. Various kinematic requirements were applied to suppress background processes such as $ \gamma p \rightarrow K^+ \Lambda/\Sigma^0$ and $\gamma p \rightarrow K^+ \Lambda^*, \Lambda^*\rightarrow \Lambda \pi^0 \gamma$. The contribution from the former can be removed by requiring a minimum missing momentum. A total of about $250~K$ events are identified as $K^+\Lambda\pi^0$ events. The latter process, however, can not be totally eliminated because of the kinematic overlapping with the reaction of interest. In Fig.~\ref{kstar} (left), the $K^+$ missing mass shows a clear $\Lambda(1405)$ and $\Lambda(1520)$ contribution when the events between the high limit of the $\pi^0$ peak and the 2~$\pi$ threshold are selected, indicating the $\Sigma^0\pi^0$ decay of the $\Lambda^*$'s. 

\section{Cross section results}
\begin{figure}[htbp]
\begin{center}
\includegraphics[width=4.5in, height=2.5in]{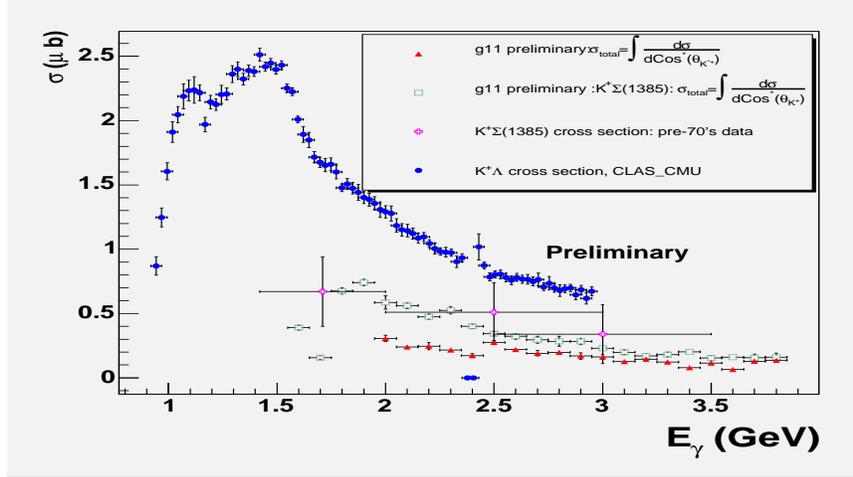}
\caption{Total cross sections of $K^*\Lambda$ and $K^+\Sigma(1385)$ compared with recent CLAS results of $K^+\Lambda$ photoproduction and the earlier measurements of $K^+\Sigma(1385)$ photoproduction.}
\end{center}
\label{crosall}
\end{figure}

The two processes of $\gamma p \rightarrow K^{*+}\Lambda$ and $\gamma p \rightarrow K^{+}\Sigma(1385)$ overlap kinematically, as shown in Fig.~\ref{kstar}(right), for events above the nominal $K^{*+}$ threshold. Extensive simulations were conducted for these two reactions assuming a t-channel process. The differential cross section results were used as the new input of the next iteration of  simulation. The $\Sigma(1385)$ yield was obtained using a $p-$wave Breit-Wigner function with background shape obtained from the simulation of $K^{*+}\Lambda$ events. The $K^{*+}$ yield was also obtained similarly. Due to the preliminary status of the results and the limit of the proceedings, only total cross section results of $K^{*+}\Lambda$ ($\sigma_{total}=\int \frac{d\sigma}{dcos^*\theta_{\Lambda}}$) and $K^{+}\Sigma(1385)$ ($\sigma_{total}=\int \frac{d\sigma}{dcos^*\theta_{K^+}}$) are included in this paper\footnote{\uppercase{T}he cross sections are preliminary, and 20\% systematic errors are expected}. Compared with the most recent CLAS results of $K^+\Lambda$ total cross section ~\cite{Bradford05} (Fig.~\ref{crosall}), it is clear that the production of $K^{*+}\Lambda$ and $K^{+}\Sigma(1385)$ is large enough that it should not be excluded from the coupled channel analysis of nucleon resonances. In the future, the comparison of the differential cross section with quark-model based calculations should also be exercised when results become available~\cite{Zhao01}.

\section{Summary and discussion}
The reaction of  $\gamma p \rightarrow K^+ \Lambda \pi^0 $ has been investigated for the  $K^{*+}\Lambda$ and $K^+\Sigma(1385)$ production. The preliminary cross section results of these two processes indicate that they are not negligible compared with $K\Lambda$ photoproduction, and should be included in the future coupled channel $N^*$ analysis. On the other hand, no significant structure was found around the 1480~MeV/c$^2$ region in the $\Lambda\pi^0$ invariant mass spectrum, in contrast with the recent results reported by COSY~\cite{ZYCHOR}.
\section{Acknowledgment}
We wish to thank all of the CLAS collaborators, the extraordinary efforts of the CEBAF staff, and particularly the g11 running group.

\end{document}